# Growing graphs with addition of communities

**E B Yudin**
Sobolev Institute of Mathematics, Siberian Branch of the Russian Academy of Sciences, 4 Acad. Koptyug avenue, 630090, Novosibirsk, Russia

The reported study was funded by RFBR, according to the research project No 16-31-60023 mol_a_dk

E-mail: udinev@asoiu.com

**Abstract.** Paper proposes a model of large networks based on a random preferential attachment graph with addition of complete subgraphs (cliques). The proposed model refers to models of random graphs following the nonlinear preferential attachment rule, and takes into account the possibility of «adding» entire communities of nodes to the network. In the derivation of the relations that determine the vertex degree distribution, the technique of finite-difference equations describing stationary states of a graph is used. The obtained results are tested empirically (by generating large graphs), special cases correspond to known mathematical relations.

## 1. Introduction

Structures studied by Network Science have different nature, but similar properties, among them the researchers distinguish the following ones:
– vertex degree in network follows a power-law distribution,
– networks have a small diameter, often decreasing with increasing network size («six handshakes» rule),
– there are communities (some groups of nodes connected to each other more than to nodes from other communities) in networks.
As models of large networks, random graphs with preferential attachment are often used [1]. Such models of growing networks allow us to generate graphs that implement:
– a given vertex degree distribution;
– a given vertex degree distribution and a given clustering coefficient [2-5];
– a given vertex degree distribution and a joint distribution of the edge end degrees [6].
It is convenient to consider the scheme of generation of preferential attachment graphs by the example of a random graph following the nonlinear preferential attachment rule (NPA graph). Generation of a NPA graph (figure 1) is an iteration process. At each step of the process an increment is added to the existing graph, an increment is a vertex with a random number of edges, whose free ends are attached to the $i$-th vertex of the existing graph with a probability proportional to the preference function $f(k_i)$, depending on the vertex degree $k_i$ of the graph $p_i = f(k_i)[\sum_{i=1}^{n} f(k_i)]^{-1}$, $i = 1,…, N$, where $N$ is the number of the vertices.

If the number of edges in the increment is a constant $m$, and the preference function is linear: $f(k_i) = k_i$, then we obtain a well-known Barabashi-Albert graph (BA) [1], that follows the principle «rich becomes richer» (the more the vertex degree, the greater probability of attaching to it). If the preference function is logarithmic: $f(k_i) = \log(k_i)$, then the law of Weber-Fechner is simulated, according to which the intensity of the perception is proportional to the logarithm the original stimulus intensity.

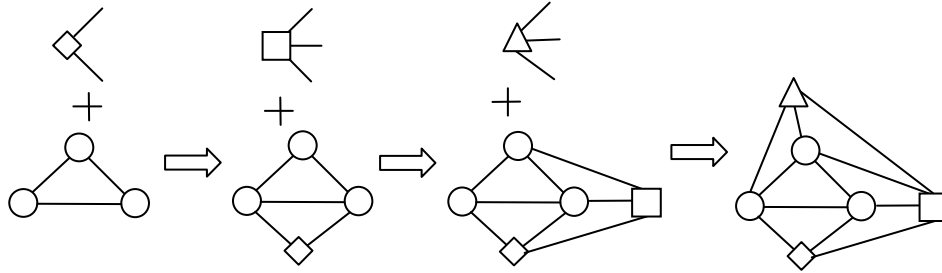

**Figure 1**. Generation of NPA graphs: vertices with a random number of edges are added to the seed graph. In the figure a vertex with two edges is added, then a vertex with three edges is added twice

However, in the graphs of preferential attachment (especially in the most studied graph of preferential attachment - the BA graph) the communities are absent as a rule, in some ways the graph is homogeneous. Even for small size complete subgraphs (cliques) containing only three vertices, in BA graphs the number of such «triangular» cliques is significantly smaller in comparison to their number in real networks with commensurate number of nodes and edges [3]. A fortiori, it is true for the number of cliques of larger size. This paper proposes preferential attachment graphs, in which this problem is solved. For these graphs it is assumed that the community is added to the network as a whole, thereby the stage of «community increment» is simulated [7]. A community increment is a given number $n$ of nodes, any two nodes of which are linked.

## 2. Theory
**Graphs with the addition of complete subgraphs.** A graph with the addition of complete subgraphs is a modified version of a NPA graph. Suppose that at every step of graph growth with a probability $\gamma$ a complete graph containing $n$ nodes ($n$-ad, i.e. polyad) is added, with the probability $(1 - \gamma)$ – the usual increment of the graph – monad (figure 2). The number of free edges of the monad is determined in accordance with the discrete distribution $\{r_k^{(1)}\}$ and for $n$-ad – in accordance with the distribution $\{r_k^{(n)}\}$. Free ends of the edges of the monad are joined to the graph independently, according to the probability

$$p_i = \frac{f(k_i)}{\sum_{j=1}^{N} f(k_j)}, \quad i = 1, \ldots, N, \tag{1}$$

where $N$ – is the number of vertices of the graph, the preference function $f(k) > 0$, if $g \leq k \leq M$, otherwise $f(k) = 0$.

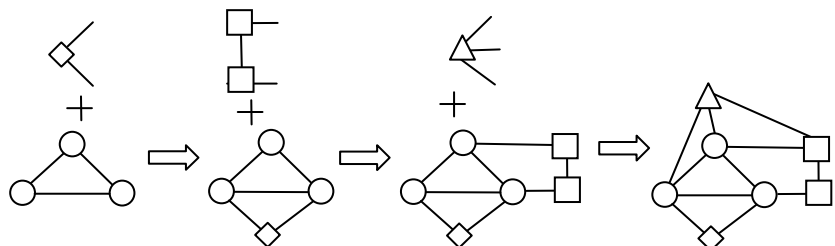

**Figure 2**. Generation of NPA graphs with communities increment, at each step of the graph generation a two-vertex increment (dyad) is added with probability $\gamma$ or a single-vertex increment (monad) is added with probability $(1-\gamma)$

Each of the μ free ends of a vertex in *n*-ad (figure 3) is attached to the same vertex, played according to the formula (1). The other free edge ends of the *n*-ad attach independently to the existing graph vertices with probability (1).

The following properties can be written for the proposed graph.

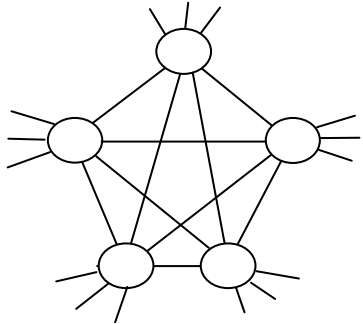

**Figure 3.** A complete graph with five vertices as an increment (pentad) with 3 played free edges and 10 included in clique.

1) On average at one step (1+(*n*–1) γ) vertices are attached.

2) Denote the average number of edges in a monad as $m^{(1)}$, then $m^{(1)} = \sum_{k=1}^{t} i \cdot r_i^1$, where *t* is the maximum value of the monad degree.

3) The average number of edges in *n*-ad (*n*·$m^{(n)}$ + *n*(*n*–1)/2), free (not included in the clique of size *n*) edges *n*·$m^{(n)}$, where $m^{(n)}$ is the average number of free edges of each vertices in the *n*-ad.

4) At one step, on average, γ·(*n*·$m^{(n)}$ + $m^{(n)}$·(*n*–1)/2) + (1–γ)·$m^{(1)}$ edges are attached to a graph.

The following tasks are set.

1) Within the framework of the proposed modified version of a NPA graph to derive equations that will allow us to calculate the vertex degree distribution (VDD) {$Q_k$} of the generated graph.

2) Within the framework of the proposed modified version of a NPA graph to derive equations that will allow us for a given VDD of real network to select generation parameters that take into account the process of communities increment.

**Derivation of the vertex degree distribution.** Let us divide the set of all vertices in the existing graph into subsets $A_k$ (layers) of vertices, containing vertices with the same vertex degree *k* ($k = m, m+1, ...$). The probability $Q_k$ of a layer $A_k$ is defined as the probability of attaching to the layer in a case of random (equiprobable) choice of any of the *N* vertices in the existing graph: $Q_k = |A_k| / N$, where $|A_k|$ is a number of vertices with the vertex degree *k*.

The final value $P_k$ of the probability $\sum_{i \in A_k} p_i$ that a vertex of the layer $A_k$ will be chosen to attach to an edge of a new vertex can be obtained from formula (1):

$$\sum_{i \in A_k} p_i = \frac{\sum_{i \in A_k} f(k_i)}{\sum_{j=1}^{N} f(k_j)} = \frac{f(k)|A_k|}{\sum_{l \geq g} f(l)|A_l|} = \frac{f_k|A_k|/N}{\sum_{l \geq g} f_l|A_l|/N} \sim \frac{Q_k f_k}{\sum_{l \geq g} Q_l f_l} = \frac{Q_k f_k}{\langle f \rangle} = P_k.$$

Thus, the probability $P_k$ of selecting a layer $A_k$ can be calculated as

$$P_k = \frac{Q_k f_k}{\langle f \rangle}, \text{ where } \langle f \rangle = \sum_{l \geq g} Q_l f_l - \text{an average value of the preference function.} \qquad (2)$$

Let us consider the factors that influence $|A_k|$ when a monad attaches.

1) If a monad attaches, then $|A_k|$ with probability $r_k$ increases by 1. On the average $|A_k|$ due to this increases by $1 + r_k^{(1)} + 0 \cdot (1 - r_k^{(1)}) = r_k^{(1)}$.

2) Moreover, on the average $m^{(1)}P_k$ monad edges attach to some vertices of the layer $A_k$ and these vertices go to the layer $A_{k+1}$, i.e. $A_k$ decreases on average by $m^{(1)}P_k$ and, for a similar reason, $A_k$ increases on average by $m^{(1)}P_{k-1}$.

In total when adding monads

$$\Delta_1 |A_k| = r_k^{(1)} + m^{(1)}P_{k-1} - m^{(1)}P_k. \qquad (3)$$

Let us consider the factors that influence $|A_k|$ when a *n*-ad (figure 4) attaches.

1) The vertices of the $n$-ad with probability $r^{(n)}_{k+1-n}$ attach to $A_k$, due to this fact $|A_k|$ increases by $n \cdot r^{(n)}_{k+1-n}$.

2) For each of the vertices in the $n$-ad it is true that $\mu$ outgoing edges attach with the probability $P_{k-n}$ to some vertices of the layer $A_{k-n}$ and these vertices go to the layer $A_k$. Due to this $|A_k|$ increases by $\mu P_{k-n}$.

3) On the average, $\mu$ pairs of conjugate edges attach to the vertices of the layer $A_k$ and these vertices go to the layer $A_{k+n}$. It gives a decrease in $|A_k|$ by the value $-\mu P_k$.

4) The free edges in the $n$-ad attach to the vertices of the layer $A_{k-1}$ with the probability $P_{k-1}$. The number of such single edges $n \cdot m^{(n)} - n \cdot \mu$. The increase in $|A_k|$ is equal $(n \cdot m^{(n)} - n \cdot \mu) P_{k-1}$.

5) The free edges in the $n$-ad attach to the vertices of the layer $A_k$ with the probability $P_k$. A decrease in $|A_k|$ is equal $(n \cdot m^{(n)} - n \cdot \mu) P_k$.

In total we obtain

$$\Delta_2 |A_k| = n \cdot r^{(n)}_{k+1-n} + \mu P_{k-n} - \mu P_k + (n \cdot m^{(n)} - n \cdot \mu) P_{k-1} - (n \cdot m^{(n)} - n \cdot \mu) P_k. \tag{4}$$

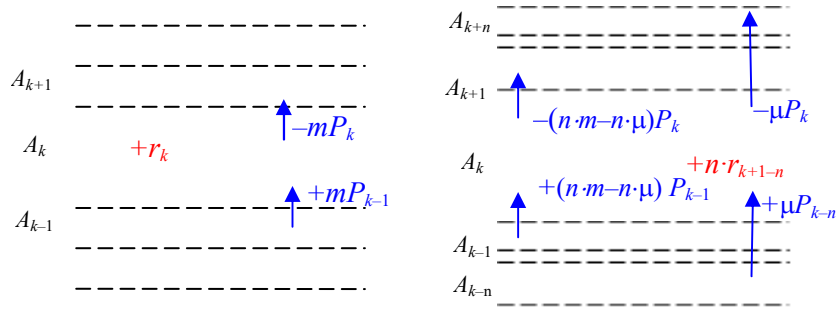

**Figure 4**. Changing the number of vertices in $A_k$ by adding of monad (left) and $n$-ad (right)

At each step of the graph growing, a module attaches with the probability $\gamma$, usual increment – with the probability $(1 - \gamma)$. In view of this

$$\Delta |A_k| = (1-\gamma) \cdot \Delta_1 |A_k| + \gamma \cdot \Delta_2 |A_k| = (1-\gamma)(r^{(1)}_k + m^{(1)} P_{k-1} - m^{(1)} P_k) +$$
$$+ \gamma \cdot (n \cdot r^{(n)}_{k+1-n} + \mu P_{k-n} - \mu P_k + (n \cdot m^{(n)} - n \cdot \mu) P_{k-1} - (n \cdot m^{(n)} - n \cdot \mu) P_k).$$

Equating the fraction of vertices in the layer $A_k$ before performing the step of graph growing to the corresponding fraction $A_{k+1}$ after this step, we obtain the following balance equation:

$$\frac{|A_k|}{N} = \frac{|A_k| + \Delta |A_k|}{N + ((1-\gamma) + \gamma \cdot n)}.$$

Or

$$\frac{|A_k|}{N} = \frac{|A_k| + r^{(1)}_k + m^{(1)} P_{k-1} - m^{(1)} P_k - \gamma \cdot r^{(1)}_k - \gamma \cdot m^{(1)} P_{k-1} + \gamma \cdot m^{(1)} P_k}{N + ((1-\gamma) + \gamma \cdot n)} +$$
$$+ \frac{\gamma \cdot n \cdot r^{(n)}_{k+1-n} + \gamma \cdot \mu \cdot P_{k-n} - \gamma \cdot \mu \cdot P_k + \gamma \cdot (n \cdot m^{(n)} - n \cdot \mu) P_{k-1} - \gamma \cdot (n \cdot m^{(n)} - n \cdot \mu) P_k}{N + ((1-\gamma) + \gamma \cdot n)}.$$

Rewrite

$$|A_k| N + |A_k| \cdot ((1-\gamma) + \gamma \cdot n) = N |A_k| + N \big( r^{(1)}_k + m^{(1)} P_{k-1} - m^{(1)} P_k - \gamma \cdot r^{(1)}_k - \gamma \cdot m^{(1)} P_{k-1} +$$
$$+ \gamma \cdot m^{(1)} P_k + \gamma \cdot n \cdot r^{(n)}_{k+1-n} + \gamma \cdot \mu \cdot P_{k-n} - \gamma \cdot \mu \cdot P_k + \gamma (n \cdot m^{(n)} - n \cdot \mu) P_{k-1} - \gamma (n \cdot m^{(n)} - n \cdot \mu) P_k \big) =$$
$$= N \cdot |A_k| + N \cdot \big( r^{(1)}_k (1-\gamma) + \gamma \cdot n \cdot r^{(n)}_{k+1-n} + (m^{(1)}(1-\gamma) + \gamma \cdot n \cdot m^{(n)} - \gamma \cdot n \cdot \mu) \cdot P_{k-1} +$$
$$\gamma \cdot \mu \cdot P_{k-n} + (m^{(1)} \cdot (\gamma - 1) - \gamma \cdot n \cdot m^{(n)} + \gamma (n-1) \cdot \mu) P_k \big).$$

Let us reduce the identical terms $|A_k|N$ and divide the equality by $N$. As a result, the expression $|A_k|((\gamma - 1) + \gamma \cdot n)$ in the left part is transformed into $Q_k((\gamma-1) + \gamma \cdot n)$, and the multipliers $N$ in the right part are reduced. Replacing all $P_k$ in the right part of equality with the expression $P_k = \dfrac{Q_k f_k}{\langle f \rangle}$, we obtain the balance equation for $Q_k$:

$$Q_k \cdot \left(1 + \gamma(n-1) - \frac{f_k(m^{(1)}(\gamma-1) - \gamma \cdot n \cdot m^{(n)} + \gamma(n-1)\mu)}{\langle f \rangle}\right) =$$

$$= \frac{(r_k^{(1)}(1-\gamma) + \gamma \cdot n \cdot r_{k+1-n}^{(n)})\langle f \rangle + (m^{(1)}(1-\gamma) + \gamma \cdot n \cdot m^{(n)} - \gamma \cdot n \cdot \mu - \gamma \cdot m^{(1)})Q_{k-1} + \gamma \cdot \mu \cdot Q_{k-n}}{\langle f \rangle}.$$

In explicit form for $Q_k$ we can write the expression:

$$Q_k = \frac{(r_k^{(1)}(1-\gamma) + \gamma \cdot n \cdot r_{k+1-n}^{(n)})\langle f \rangle + (m^{(1)}(1-\gamma) + \gamma \cdot n \cdot m^{(n)} - \gamma \cdot n \cdot \mu)f_{k-1}Q_{k-1} + \gamma \cdot \mu \cdot f_{k-n}Q_{k-n}}{\langle f \rangle + \langle f \rangle \gamma(n-1) + f_k[m^{(1)} \cdot (1-\gamma) + \gamma \cdot n \cdot m^{(n)} - \gamma(n-1) \cdot \mu]}, \quad (5)$$

where $k = g, g+1, ..., M$, $\langle f \rangle = \sum_{k \geq g} f_k Q_k$.

If there is no $n$-ad, i.e. $\gamma = 0$, then we obtain a well-known formula [8]:

$$Q_k = \frac{r_k^{(1)}\langle f \rangle + m^{(1)} \cdot f_{k-1}Q_{k-1}}{\langle f \rangle + m^{(1)} \cdot f_k}.$$

If there is no monad, i.e. $\gamma = 1$, at $n = 2$ we obtain:

$$Q_k = \frac{2 \cdot r_{k-1}^{(2)}\langle f \rangle + (2m^{(n)} - 2\mu) \cdot f_{k-1}Q_{k-1} + \mu \cdot f_{k-2}Q_{k-2}}{2\langle f \rangle + 2f_k \cdot m^{(2)} - f_k \mu}.$$

**Calibration of the graph by the vertex degree distribution.** In the problem of graph calibration it is necessary to find the parameters $\{r_k^{(1)}\}, \{r_k^{(n)}\}, \{f_k\}, \gamma, \mu$ at which the graph will have a given VDD $\{Q_k\}$. This problem can be solved using equation (5).
Expressing $f_k$ from (5) we obtain the recurrence relation

$$f_k = \frac{[r_k^{(1)}(1-\gamma) + \gamma \cdot n \cdot r_{k+1-n}^{(n)}]\langle f \rangle}{a \cdot Q_k} - \frac{[\langle f \rangle + \langle f \rangle \gamma(n-1)]Q_k}{a \cdot Q_k} + $$
$$+ \frac{[(m^{(1)}(1-\gamma) + \gamma \cdot n \cdot m^{(n)} - \gamma \cdot n \cdot \mu)f_{k-1}Q_{k-1} + \gamma \mu f_{k-n}Q_{k-n}]\langle f \rangle}{\langle f \rangle \cdot a \cdot Q_k}, \quad (6)$$

where $a = m^{(1)} \cdot (1-\gamma) + \gamma \cdot n \cdot m^{(n)} - \gamma(n-1) \cdot \mu$.
Relation (6) defines a sequence $\{f_k\}$ defining the preference function, each term of the sequence has a common multiplicative factor $\dfrac{\langle f \rangle}{a}$. It should be noted that the sequences $\{f_k\}$ and $\{Cf_k\}$, which differ by the factor $C > 0$, are equivalent, taking into account the formula (1). Therefore, we can equate the factor $\dfrac{\langle f \rangle}{a}$ on the right side of the relation (6) to one (thus determining the equality $\langle f \rangle = a$) and write this ratio in a more compact form:

$$f_k = \frac{r_k^{(1)}(1-\gamma) + \gamma n r_{k+1-n}^{(n)} - (1+\gamma(n-1))Q_k}{Q_k} + \frac{(m^{(1)}(1-\gamma) + \gamma n m^{(n)} - \gamma n \mu)f_{k-1}Q_{k-1} + \gamma \mu f_{k-n}Q_{k-n}}{\langle f \rangle \cdot Q_k}. \quad (7)$$

Taking into account that as a result of this choice we have equality $\langle f \rangle = a$, and the parameter $a = m^{(1)} \cdot (1-\gamma) + \gamma \cdot n \cdot m^{(n)} - \gamma(n-1) \cdot \mu$, we obtain the following recurrence relation

$$f_k = \frac{r_k^{(1)}(1-\gamma) + \gamma \cdot n \cdot r_{k+1-n}^{(n)} - (1+\gamma(n-1))Q_k}{Q_k} + \frac{(m^{(1)}(1-\gamma) + \gamma \cdot n \cdot m^{(n)} - \gamma \cdot n \cdot \mu)f_{k-1}Q_{k-1} + \gamma \mu f_{k-n}Q_{k-n}}{[m^{(1)} \cdot (1-\gamma) + \gamma \cdot n \cdot m^{(n)} - \gamma(n-1) \cdot \mu] \cdot Q_k}, \quad (8)$$

where $k = g, g+1, ..., M$.

Formula (8) is the solution of graph calibration problem. To realize a given VDD $\{Q_k\}$ it allows to obtain infinitely many weights sequences $\{f_k\}$, which actually set the preference function $f(k)$, depending on which parameters $\{r_k\}$, $\gamma$, $\mu$ and $m$ are chosen.

## 3. Numerical experiments

Figure 5 shows a graph calibration using pentads (complete graphs in increments that contain 5 vertices) and monads. The probability of pentad occurrence $\gamma = 0.01$. As a basic vertex degree distribution we get a smoothed distribution, obtained from data on the network of autonomous Internet systems (available at http://www-personal.umich.edu/~mejn/netdata/as-22july06.zip). Figure 5 shows the results of generating a graph with given parameters after adding 50 thousand increments, a complete graph containing 4 vertices was used as a seed graph.

Figure 5. Calibration of a random graph using spreadsheets

## 4. Conclusion

Despite the numerous publications on the analysis of communities and their dynamics in real networks, the number of papers solving more fundamental questions such as mechanisms of communities emergence modeling is few. For example, in [9, 10] geometric graphs with preferential attachment are proposed (where the geometry determines the degree of similarity, affecting the preference of attachment and defining the form of communities), papers [11, 12] propose to "generate" graphs as separate weakly connected components (they can also be considered as communities). This paper proposes a random graph that simulates the addition of communities as a whole. This process interpretation can be found in social networks. Thus, when the next study group appears, new nodes (representing students) are added to the subnet, the nodes immediately become interconnected (students usually communicate in the subnet with other students from their study group).

Accounting other processes that take place in networks can be referred as a further development of the proposed model, it can be effectively performed with the help of agent-based modeling [13, 14]. By analyzing the dynamics of changes in the network structure papers [15, 16] reveal such processes as destruction of communities (division into several smaller ones or complete destruction of the community), and connection between the age of the community and its size.

**Acknowledgments**

The reported study was funded by RFBR, according to the research project No 16-31-60023 mol_a_dk.